Article type:    Original research

Title:           Gravitational anti-screening as an alternative to the ΛCDM model

Author:          A. Raymond Penner

Address:         Department of Physics,
                 Vancouver Island University,
                 900 Fifth Street,
                 Nanaimo, BC, Canada,
                 V9R 5S5

Email:           raymond.penner@viu.ca

Tel:             250 753-3245 ext: 2336

Fax:             250 740-6482

**Gravitational anti-screening as an alternative to the ΛCDM model**

**Abstract**

Previously, in Penner (2016), a theory of gravitational anti-screening was shown to lead to the Baryonic Tully-Fisher Relationship. In addition it was shown to agree with the observed rotation curve of the Galaxy, the observed features in the rotational curves of other spiral galaxies, and with observations of the Coma cluster. In this paper, the theory is now shown to be consistent with a geometrically flat universe. Using a model of the distribution of superclusters, the overall density parameter of the universe, as determined by the theory, is $\Omega = 1.08 \pm 0.19$. In addition, the energy density which falls out from the theory has a negative pressure associated with it. This, along with a model of the evolution of superclusters, leads to an acceleration of the universal expansion without the requirement of dark energy. The theory of gravitational anti-screening therefore provides an alternative to the ΛCDM model of cosmology.

# 1. Introduction

In the ΛCDM model of cosmology, the three primary constituents of the universe are ordinary baryonic matter, dark matter, and dark energy. Dark matter is usually taken to consist of currently undetected non-baryonic particles with the prime candidate being stable weakly interacting massive particles (WIMPs) predicted by supersymmetry models. Dark energy is, most commonly, taken to be a constant vacuum energy density that fills all space and which has a negative pressure associated with it. The relative contributions that each of these three constituents makes to the overall energy density of the universe stems primarily from the following observations:

i)      Measurements of CBR anisotropies by Ade et al (2014) have determined that the density parameter for baryons, $\Omega_{B,O}$, is given by

$$\Omega_{B,O} = 0.0482 \pm 0.0016 \qquad\qquad (1)$$

for their determined Hubble constant of

$$H_O = (67.80 \pm 0.77) \text{ km s}^{-1} \text{ Mpc}^{-1}. \qquad\qquad (2)$$

Models of BBN and observations of the abundance of the light elements are in agreement with this value (Burles et al 2001, Kirkman et al 2003).

ii)      The existence of dark matter is inferred from observations of the rotational velocities of stars and gas clouds in the outer regions of spiral galaxies. These velocities are found to be much greater than what is predicted from the baryonic mass of these galaxies. Examples of galactic rotational curves are provided by Sofue (1996), Sofue et al (1999) and Nordermeer et al (2007). It is difficult to determine exactly how much additional mass is needed. However, it would appear that an order of magnitude more mass than what is provided by the baryonic mass is required. The velocity dispersions of galaxies in galactic clusters also indicate that approximately an order of magnitude more mass than what is provided by the baryonic mass is required to keep the clusters bound (Zwicky 1933, Lokas&Mamon 2003, Merritt 1987).

iii)    The inclusion of dark energy stems from observations by Perlmutter et al (1999) and Riess et al (1998) of the light curves and redshifts of type Ia supernovae at high z which indicate that the expansion of the universe is accelerating.

iv)    Measurements of CBR anisotropies by Hinshaw et al (2013) have determined that the universe's overall density parameter is given by

$$\Omega = 1.0027 \pm 0.0039 \tag{3}$$

Other theoretical considerations also seem to indicate that $\Omega$ must be almost exactly equal to 1.

A best fit provided by Ade et al (2014) of the ΛCDM model to observations i, iii, and iv leads to the following contributions that the three constituents make to the overall density parameter;

| | | |
|---|---|---|
| Baryonic matter: | $\Omega_{B,O} = 0.0482,$ | (4a) |
| Dark matter: | $\Omega_{DM,O} = 0.2586,$ | (4b) |
| Dark energy: | $\Omega_{\Lambda,O} = 0.692.$ | (4c) |

For these values, the deceleration parameter, $q_O$, with the equation of state parameter, $w_i$, for both baryonic and dark matter being set equal to 0 and for dark energy being set equal to -1, is given by;

$$q_O = \frac{1}{2} \sum \Omega_{i,O}(1+3w_i) = \text{-0.539.} \tag{5}$$

Alternatives to the ΛCDM model are primarily related to the dark matter component and its ability to explain details of the rotational velocities of galaxies. Observations indicate that these rotational velocities approach a constant value in the outer regions with little indication that they will eventually fall off in a Keplerian fashion. This is in disagreement with the standard dark matter model. Indeed, related to this, there is a primary observation for which the standard dark matter model does not provide a natural explanation;

v)    An empirical relationship exists between $M_B$, the baryonic mass of a galaxy, and v, the galaxy's constant outer rotational velocity. This is referred to as the Baryonic Tully-Fisher Relationship (BTFR) (McGaugh et al 2000, McGaugh 2012). The BTFR as given by McGaugh (2012) is

$$M_B = A\ v^4 \tag{6a}$$

with

$$A = (47 \pm 6) \, M_\odot \, km^{-4} \, s^4. \tag{6b}$$

The BTFR can also be expressed as the following relationship between M, the apparent mass of the galaxy, its baryonic mass, and the observation distance r, by substituting $GM/r$ for $v^2$ in (6a);

$$M = (G^2 A)^{-1/2} r \, M_B{}^{1/2}. \tag{7}$$

The BTFR is a relationship with surprisingly little scatter that ranges over five orders of magnitude of galactic baryonic masses.

The BTFR is the cornerstone of the alternatives to the dark matter component of the ΛCDM model. Some of these alternatives that have been put forward (Blanchet 2007a, 2007b; Blanchet and LeTiec 2008; Penner 2011, 2012; Hajdukovic 2011a, 2011b, 2011c, 2012 a, 2012b) are based on the hypothesis that mass dipoles exist throughout the cosmos. These dipoles become gravitationally polarized in the presence of an external gravitational field which leads to an anti-screening of the baryonic mass. These various models differ as to what the mass dipoles actually are and how they behave in a gravitational field. In Penner (2016) an overview of these mass dipole models is provided.

In Penner (2016) the author looked at the possibility of mass dipoles and the anti-screening of a baryonic mass from a different perspective. The case where the polarization of the vacuum does occur, namely in electrodynamics, was modeled. This model was then carried over to the gravitational case. In addition to naturally leading to the BTFR, this gravitational anti-screening model was shown to be in good agreement with the determined rotational curve of the Galaxy and the velocity dispersions and shear values for the Coma cluster.

Although, the gravitational anti-screening theory presented in Penner (2016) does not preclude dark energy, this paper will demonstrate that dark energy is no longer required. By applying the model from Penner (2016) to the major baryonic mass component of the universe, i.e. superclusters, it will be shown that the theory of gravitational anti-screening is in agreement with an overall density parameter equal to 1. In addition, the energy density which falls out from the theory has a negative pressure associated with it. This, along with a model of the evolution of superclusters, leads to an acceleration of the universal expansion.

## 2. Theory and Results

## 2.1 Gravitational anti-screening

As indicated above the model of gravitational anti-screening presented in Penner (2016) was obtained by considering the case of the screening that is found in quantum electrodynamics. In quantum electrodynamics, a charged particle is taken to be surrounded by a cloud of virtual photons which spend part of their existence dissociated into pairs of virtual particle-antiparticles. These particle-antiparticle pairs have an electric dipole moment with no net charge. As with the classical model of a point charge within a dielectric, the virtual electric dipole pairs within the vacuum will provide a screening effect so that the apparent charge of the particle is less than its true value. As discussed in Penner (2016), to get this behavior, $\mathbf{P_E}$, the virtual electric dipole moment density, must have the following dependence on $\mathbf{E}$, the electric field strength;

$$\mathbf{P_E} \rightarrow \chi \epsilon_o \mathbf{E} \quad \text{as } E \rightarrow 0 \qquad \text{and} \tag{8a}$$

$$\mathbf{P_E} < \chi \epsilon_o \mathbf{E} \quad \text{as } E \rightarrow \infty. \tag{8b}$$

where $\chi$ is a dimensionless constant known in the classical case as the electric susceptibility of a dielectric. The specific function chosen to model this behavior is found to have only a secondary effect. In Penner (2016), the following function, which has the behavior given by (8), was used

$$\mathbf{P_E} = \chi \epsilon_o E_o \left( 1 - e^{-E/E_o} \right) \hat{\mathbf{E}} \tag{9}$$

where $E_o$ is a constant to be determined. For this modeled behavior the apparent charge of an object falls rapidly with distance to a constant value which is $1/1+\chi$ of its true value.

The dependence that the virtual electric dipole moment density has on the electric field, as given by (9), was then applied in Penner (2016) to the hypothetical case where like charges attract and unlike charges repel. The virtual entities in the vacuum then would provide an anti-screening effect resulting in the apparent charge being greater than its true value. However, in this case the apparent charge does not approach a constant value but increases with observational distance. For the specific case where $\chi = -1$ and at large observational distances, the dependence that Q, the apparent charge, has on r, the observational distance, and $Q_O$, the true charge, was found to be given by;

$$Q = (8\pi\epsilon_o E_o)^{1/2} \, r \, Q_O^{1/2}. \tag{10}$$

The agreement between (7) and (10) led the author to propose the following hypothesis (Penner 2016):

Particles and thereby baryonic masses are surrounded by a cloud of virtual gravitons which spend part of their existence dissociated into virtual entities. The virtual entities have no net mass but do have a mass dipole moment. In addition, the dependence that the virtual mass dipole moment density has on the gravitational field is the same as the modeled dependence that the virtual electric dipole moment density has on the electric field.

The exact nature of the hypothesized entities was not speculated on

By the above hypothesis the dependence that, $\mathbf{P_G}$, the virtual mass dipole density, surrounding a given baryonic mass has on $\mathbf{g}$, the total gravitational field is given by (Penner 2016);

$$\mathbf{P_G} = -\frac{1}{4\pi G} \, g_o \, (1 - e^{-g/g_o}) \, \hat{\mathbf{g}} \, , \tag{11}$$

where $g_o$ is a model parameter to be determined empirically. The resulting energy density of the vacuum, $\varepsilon_v$, surrounding any baryonic mass will then be given by;

$$\varepsilon_v = (-\boldsymbol{\nabla} \cdot \mathbf{P_G}) \, c^2 \tag{12}$$

This energy density provides the additional gravitational field that is currently attributed to dark matter. As shown in Penner (2016), this model leads to the following far field relationship between M, the apparent mass, $M_B$, the baryonic mass, and r, the observation distance;

$$M = \left(\frac{2g_o}{G}\right)^{1/2} r \, M_B^{1/2}. \tag{13}$$

Equation (13) is of course the BTFR. The value of the parameter $g_o$ was then obtained by equating the coefficients of (7) and (13);

$$g_o = \frac{1}{2GA} = (8.0 \pm 1.0) \times 10^{-11} \text{ m s}^{-2}. \tag{14}$$

This model was then used to determine the rotational curve of the Galaxy and the velocity dispersions and shear values for the Coma cluster. These results were found to be in good agreement with observations.

## 2.2 Model of supercluster distribution

Observations show that the galaxies in the universe are not uniformly distributed but are typically found in groups and clusters. These groups and clusters are in turn part of superclusters separated by large sparsely populated voids. From a catalog of superclusters out to $z \leq 0.08$, it is estimated that the mean separation between superclusters is $D_{SC} \cong 100h^{-1}$ Mpc (Bahcall&Soneira 1984, Bahcall 1996). Observations by Broadhurst et al (1988) of the redshift distribution of galaxies in narrow pencil-beam surveys in turn indicate an apparent periodic distribution of galaxies with a regular separation of $128h^{-1}$ Mpc. This distribution was later shown by Bahcall (1991) to originate from the intersection of the narrow-beams with the tails of large superclusters. Distances between high-density regions across the voids were also determined using narrow pencil beam surveys (Einasto et al 1997). The median distances for the different samples ranges from $116h^{-1}$ Mpc to $143h^{-1}$ Mpc. From these listed results the current average separation of superclusters is estimated to be

$$D_{SC,O} = (120 \pm 20)h^{-1} \text{ Mpc}. \tag{15a}$$

and for the Hubble constant as given by (1)

$$D_{SC,O} = (177 \pm 32) \text{ Mpc}. \tag{15b}$$

This separation is also approximately equal to the diameter of an average void (Tully 1986, Batuski&Burns 1985, Einasto et al 1997).

As a simplified model of the baryonic mass distribution of the universe, it will be taken that the baryonic mass is lumped together at the location of uniform sized superclusters currently separated by $D_{SC,O}$. Each model supercluster will be taken to be the dominant source of the gravitational field for distances within $R_{SC,O}$ of the superclusters centre where

$$R_{SC,O} = D_{SC,O}/2 = (89 \pm 16) \text{ Mpc}. \tag{16}$$

The total baryonic mass associated with these modeled superclusters will be such that the average baryonic density within $R_{SC,O}$ is as given by (1). This leads to each modeled supercluster having a total baryonic mass of approximately $3 \times 10^{16} M_{\odot}$. Although this supercluster model is very simple, the dominant contribution that the polarized vacuum will make to the overall energy density of the universe will come from the voids. It is the behavior and values of the gravitational fields within the voids that is of greatest importance. In this case, the above model of baryonic mass would result in gravitational fields within the voids that would be expected to be in reasonable agreement with actual values.

### 2.3 Density parameter

With the given model the total mass, $M$, within a distance of $R_{SC}$ from a supercluster with a baryonic mass of $M_B$ is from (13) given by

$$M = \left(\frac{2g_0}{G}\right)^{1/2} R_{SC} \, M_B^{1/2}. \tag{17}$$

Eq. (17) can also be expressed in terms of $\varepsilon_B$, the average baryonic density, and $\varepsilon$, the average overall energy density, i.e.

$$\varepsilon = \left(\frac{3g_0 c^2}{2\pi G}\right)^{1/2} \left(\frac{\varepsilon_B}{R_{sc}}\right)^{1/2}. \tag{18}$$

In terms of density parameters (18) becomes

$$\Omega = \left(\frac{4g_0}{H^2}\right)^{1/2} \left(\frac{\Omega_B}{R_{sc}}\right)^{1/2}. \tag{19}$$

Substituting (1), (2), (14), and (16) into (19), the current value of the overall density parameter as given by the theory of gravitational anti-screening is

$$\Omega = 1.08 \pm 0.19. \tag{20}$$

Given the simplicity of the model of superclusters the result is quite good. Their baryonic mass plus the contribution from the polarized vacuum is consistent with $\Omega$ being equal to 1. No additional contributor to the energy content of the universe, such as dark energy, is required to explain observation iv.

## 2.3 Deceleration parameter

Of course, the primary reason for invoking dark energy was to explain the acceleration of the expansion of the universe. As stated in Section 1, a best fit of the ΛCDM model by Ade et al (2014) leads to a deceleration parameter $q_O$ is equal to -0.539. In the author's gravitational anti-screening theory the baryonic mass and polarized vacuum are coupled together with a total energy density given by (18). Substituting

$$R_{SC} = a\ R_{SC,O} \tag{21}$$

and

$$\varepsilon_B = \varepsilon_{B,O}\ a^{-3} \tag{22}$$

into (18) results in the following dependence that the overall average energy density has on the scale factor;

$$\varepsilon = \left(\frac{3g_0 c^2}{2\pi G}\right)^{1/2} \left(\frac{\varepsilon_{B,O}}{R_{sc,O}}\right)^{1/2} a^{-2}. \tag{23}$$

Given this dependence on scale factor the equation of state parameter for the model is $w = -1/3$ which, as with dark energy, leads to a negative pressure. By equation (5) the deceleration parameter for this single component model will therefore equal 0 and the universe will expand at constant velocity. From the Friedmann equation, à, the velocity of this expansion, is equal to $H_O$.

The model of the baryonic mass distribution of the universe presented assumes that the baryonic distribution of the universe is static, i.e. as the universe expands the superclusters are just moving farther apart. If superclusters are, in general, at least partially unbound systems then the baryonic mass of the superclusters would be expected to decrease over time i.e. the universal expansion is pulling the gravitationally unbound clusters away. As an example of the effect that this would have on the deceleration parameter, consider the simple case where $R_{SC}$ stays constant in time with the baryonic mass of the individual superclusters decreasing such that the average baryonic energy density continues to follow (22). This will correspond to the size of the voids remaining constant as the number of modeled superclusters increases with time. For the case of a constant $R_{SC}$, by (22) and (18), the dependence that the overall average energy density has on the scale factor is given by

$$\varepsilon = \left(\frac{3g_0 c^2}{2\pi G}\right)^{1/2} \left(\frac{\varepsilon_{B,O}}{R_{sc}}\right)^{1/2} a^{-3/2} \tag{24}$$

For this dependence on scale factor the equation of state parameter is $w = -1/2$, resulting in a deceleration parameter of $q_O = -0.25$. The magnitude of the deceleration parameter for this example is smaller than the observed value, but given the simplicity of the model the difference is not surprising. In general, if the baryonic mass distribution of the universe is static then the deceleration parameter is equal to 0 while if the baryonic mass is dispersing the deceleration parameter will be negative. The key point is that an accelerating expansion can fall out of the theory of gravitational anti-screening without the need of dark energy.

## 3. Conclusion

The theory of gravitational anti-screening provides an alternative to the standard ΛCDM model. Previously, in Penner (2016), the theory was shown to lead to the BTFR, and to agree with the observed rotation curve of the Galaxy, the observed features in the rotational curves of other spiral galaxies, and with observations of the Coma cluster. In this paper the theory is now shown to be consistent with a geometrically flat universe. Using a model of the distribution of superclusters, the overall density parameter of universe, as given by the theory, is determined to be $\Omega = 1.08 \pm 0.19$. In addition, the energy density which falls out from the theory has a negative pressure associated with it. This, along with a model of the evolution of superclusters, leads to an acceleration of the universal expansion without the requirement of dark energy.

The author's theory is a variation of the standard dark matter model in the sense that it is an energy density that contributes to the gravitational field surrounding a given baryonic mass distribution. No changes in gravitational theory or Newtonian mechanics are required. However, unlike the standard model of dark matter, the additional contribution provided is directly coupled to the baryonic mass. The author's theory may also be seen as a variation of the standard dark energy model since the polarized vacuum provides a positive energy contribution with an associated negative pressure. However, unlike the standard model of dark energy, the vacuum contribution is again directly coupled to the baryonic mass.

One feature that is especially appealing about the theory is the role of the baryonic mass. The induced vacuum energy contributes to the gravitational field surrounding a baryonic mass and to the energy content of the universe, but this contribution is solely dependent on the baryonic mass distribution; a universe without baryonic mass would truly be empty. Astronomical observations that lead to the determination of the baryonic mass distribution will lead directly to the determination of the dynamics of galaxies, clusters, superclusters, and the expansion of the universe as a whole. It is important to stress that the theory of gravitational anti-screening has no free parameters. The only parameter in the model is $g_o$ which is determined empirically from the coefficient of the BTFR. This gives the theory superior predictive powers when compared to the ΛCDM cosmological model.

Future work will apply this theory of gravitational anti-screening to binary galaxies and to the solar system. It is expected that analysis of binary galaxies will provide the best differentiation between the standard dark matter model and the theory of gravitational anti-screening while the solar system will be the best laboratory for making predictions from the theory that can be tested.